\UseRawInputEncoding
\documentclass[twocolumn,final,aps,prl,showpacs,superscriptaddress,amsmath,amssymb,amsfonts,floatfix]{revtex4-1}

\usepackage{graphicx}
\usepackage{multirow}
\usepackage{epsfig}
\usepackage{url}
\usepackage{color}

\usepackage[colorlinks,breaklinks,bookmarks=true,citecolor=blue,linkcolor=blue,urlcolor=blue]{hyperref}
\usepackage{color}
\usepackage{tabularx}
\usepackage{sidecap}
\usepackage{soul}
\usepackage{float}

\usepackage[colorlinks,breaklinks,bookmarks=true,citecolor=blue,linkcolor=blue,urlcolor=blue]{hyperref}
\usepackage{color}
\usepackage{orcidlink}

\begin{document}

\author{Ying Gao\,\orcidlink{0009-0006-4153-5193}}
\thanks{These authors contributed equally.}
\affiliation{School of Physics, Northwest University, Xi'an 710127, China}

\author{Wenfeng Wu\,\orcidlink{0000-0002-6575-5813}}
\thanks{These authors contributed equally.}
\affiliation{Key Laboratory of Materials Physics, Institute of Solid State Physics, HFIPS, Chinese Academy of Sciences, Hefei 230031, China}
\affiliation{Science Island Branch of Graduate School, University of Science and Technology of China, Hefei 230026, China}
\affiliation{Institute of Solid State Physics, TU Wien, 1040 Vienna, Austria}

\author{Zhaoxin Liu}
\affiliation{School of Physics, Northwest University, Xi'an 710127, China}

\author{Karsten Held\,\orcidlink{0000-0001-5984-8549}}
\email{held@ifp.tuwien.ac.at}
\affiliation{Institute of Solid State Physics, TU Wien, 1040 Vienna, Austria}

\author{Liang Si\,\orcidlink{0000-0003-4709-6882}}
\email{siliang@nwu.edu.cn}
\affiliation{School of Physics, Northwest University, Xi'an 710127, China}
\affiliation{Shaanxi Key Laboratory for Theoretical Physics Frontiers, Xi'an 710127, China}
\affiliation{Institute of Solid State Physics, TU Wien, 1040 Vienna, Austria}

%\title{Proton Evolution Induces Single-band $d$-wave Superconductivity in La$_2$NiO$_4$}
\title{Topotactical Hydrogen Induced Single-Band $d$-Wave Superconductivity in La$_2$NiO$_4$}

\begin{abstract} 
La$_2$NiO$_4$ is an antiferromagnetic insulator with a structural resemblance to its cuprate counterpart, La$_2$CuO$_4$. However, La$_2$CuO$_4$ 
has a Cu$^{2+}$ or 3$d^9$ electronic configuration that needs to be hole or electron doped for superconductivity, whereas La$_2$NiO$_4$  is  3$d^8$ with divalent Ni$^{2+}$.
Making a cuprate analog through conventional electron doping is impractical due to the rarity of tetravalent substituents for trivalent La.
Here, we propose an alternative route:
intercalating topotactical hydrogen, which is possible through electric-field-controlled protonation and transforms La$_2$NiO$_4$ into a 3$d_{x^2-y^2}$ single-band two-dimensional antiferromagnetic Mott insulator analogous to La$_2$CuO$_4$. This we find through density-functional theory and dynamical mean-field theory calculations.
The furthergoing dynamical vertex approximation  predicts that H-La$_2$NiO$_4$ can host $d$-wave superconductivity under 15\% hole doping with a critical temperature above 20\,K. Our findings not only suggest a new method for tuning the electronic structure of layered nickelates but also provide theoretical evidence for a new nickelate superconductor, awaiting experimental synthesis.
\end{abstract}

\maketitle

\emph{Introduction---}The discovery of new superconductors \cite{norman2016materials} is one of the central goals in condensed matter physics, as each breakthrough not only deepens our fundamental understanding of the pairing mechanism but also paves the way toward achieving room-temperature superconductivity.
Among the various families of superconductors, the cuprates \cite{bednorz1986possible,anderson1997theory,RevModPhys.72.969,RevModPhys.75.473}---characterized by a single-band Fermi surface originating
from $d$-$p$ hybridization---hold the record for unconventional superconductivity at ambient pressure with critical temperatures exceeding 130\,K \cite{schilling1993superconductivity}. This remarkable performance has motivated extensive efforts to identify new superconducting systems that share similar electronic characteristics with cuprates.

In recent years, the quest for cupratelike superconductivity has extended to nickelates \cite{allan1989comparison,bates1989normal,acrivos1994paramagnetism,PhysRevB.70.165109,PhysRevB.59.7901,PhysRevLett.100.016404,PhysRevLett.103.016401,guo2018antiferromagnetic,zhou2018observation}. Early attempts in this direction were  unsuccessful until 2019, when Li and Hwang \emph{et al.}~realized superconductivity in Sr-doped NdNiO$_2$ via a reduction process from (Nd,Sr)NiO$_3$ by CaH$_2$ \cite{li2019superconductivity,PhysRevLett.125.027001}. This breakthrough subsequently led to the discovery of a range of superconducting nickelates, including infinite-layer (La,Ca)NiO$_2$ \cite{zeng2022superconductivity}, (La,Sr)NiO$_2$ \cite{osada2021nickelate}, (Pr,Sr)NiO$_2$ \cite{osada2020phase}, (Nd,Ca)NiO$_2$, finite-layer Nd$_6$Ni$_5$O$_{12}$ \cite{pan2022superconductivity,yan2025superconductivity}, pressurized (La,Pr)$_3$Ni$_2$O$_7$ \cite{sun2023signatures,hou2023emergence,wang2024bulk}, pressurized (La,Pr)$_4$Ni$_3$O$_{10}$ \cite{zhu2024superconductivity,li2024signature,pei2024pressure,zhang2025bulk,huang2024signature}, and La$_3$Ni$_2$O$_7$ under ambient pressure and compressive strain \cite{ko2024signatures,zhou2024ambient,li2025photoemission}. These developments have restarted the search for nickelate and even palladate superconductors \cite{PhysRevLett.130.166002} that  mirror the single-band physics observed in cuprates.

Although various theoretical models have been proposed for both infinite-layer \cite{kitatani2020nickelate,PhysRevB.101.081110,PhysRevX.10.041002,PhysRevB.101.020501,PhysRevB.101.041104,PhysRevResearch.2.013214,PhysRevLett.125.077003,PhysRevResearch.2.023112,PhysRevB.100.205138,zhou2020spin,PhysRevB.106.035111,PhysRevLett.133.126401} and finite-layer nickelate \cite{PhysRevMaterials.6.L091801,PhysRevB.105.155109,PhysRevB.105.085118,PhysRevB.106.115132,PhysRevB.110.235107,PhysRevResearch.4.043036} superconductors---ranging from single-band to multiband descriptions---recent THz experiments \cite{cheng2024evidence,PhysRevB.111.014519} and resonant inelastic x-ray scattering (RIXS) measurements \cite{PhysRevB.109.235126} indicate that infinite-layer  nickelate  superconductors exhibit an electronic structure with a doped Hubbard single band of $d_{x^2-y^2}$ character similar to that of two-dimensional cuprate superconductors, suggesting $d$-wave superconductivity.
For the pressurized (La,Pr)$_3$Ni$_2$O$_7$ \cite{sun2023signatures,hou2023emergence,wang2024bulk} and pressurized (La,Pr)$_4$Ni$_3$O$_{10}$ \cite{zhu2024superconductivity,li2024signature,pei2024pressure,zhang2025bulk,huang2024signature} systems, the incorporation of an apical oxygen atom leads to Ni-3$d_{z^2}$ orbital contributions near the Fermi surface. Theoretical studies suggest that the superconductivity in these pressured samples may result either from a Zhang-Rice-like $d$-$p$ hybridization between 
(in-plane) O-$p_x$/$p_y$ and Ni-3$d_{x^2-y^2}$ \cite{sun2023signatures}, or $s^{\pm}$-wave pairing \cite{zhang2024structural,PhysRevB.109.L220506,PhysRevB.108.165141,PhysRevLett.133.136001,PhysRevB.108.L140505,PhysRevB.110.L180501,PhysRevB.109.L220506,PhysRevLett.131.236002,PhysRevB.110.L180501} induced by a charge or spin density wave  \cite{ni2025spin,PhysRevMaterials.8.L111801,li2025distinct,PhysRevLett.132.256503,PhysRevX.13.041018,PhysRevMaterials.8.L111801,PhysRevB.110.L140508,meng2024density,kumar2020physical,liu2024electronic,PhysRevB.109.235123}.

Among various nickelates, La$_2$NiO$_4$  has long attracted interest due to its structural similarity to La$_2$CuO$_4$ \cite{allan1989comparison,acrivos1994paramagnetism,PhysRevB.40.4463,PhysRevB.40.2229,brown1992modelling,rodriguez1991neutron,PhysRevB.110.205110,bialo2024strain,cui2024strain}, the parent compound of the cuprate superconductors. While the undoped La$_2$NiO$_4$ is an antiferromagnetic Mott insulator exhibiting intriguing quantum states \cite{guo1988electronic,rodriguez1991neutron,PhysRevB.40.4463,PhysRevB.55.9203}, including charge and orbital ordering \cite{PhysRevB.75.045128,PhysRevB.70.064105,PhysRevB.59.3775,PhysRevB.64.165106,PhysRevB.55.9203}---especially under hole doping---the Ni ions in La$_2$NiO$_4$ stay in a Ni$^{2+}$ state or 3$d^8$ electronic configuration. Achieving superconductivity akin to that in cuprates requires an effective 3$d^{9-\delta}$ ($\delta$ $\sim$ 0.2) electronic configuration \cite{anderson1997theory,RevModPhys.72.969,scalapino1995case}, which, in principle, demands excessive electron doping. However, electron doping in La$_2$NiO$_4$ is impractical due to the absence of tetravalent elements capable of substituting La.

An alternative and promising route is proton evolution which is realized by applying ionic liquid gating and which recently has proven effective in controlling the electronic and magnetic properties of correlated oxides such as SrRuO$_3$ \cite{li2020reversible}, SrCoO$_{2.5}$ \cite{li2019electric,lu2022enhanced}, CaRuO$_3$ \cite{PhysRevX.11.021018}, and Ca$_2$RuO$_4$ \cite{PhysRevMaterials.8.074408} that are structurally analogous to La$_2$NiO$_4$. In these systems, the intercalated proton acts as an electron donor by forming (OH)$^-$. Motivated by these advances of liquid gating, we investigate the effect of proton evolution in La$_2$NiO$_4$.

To this end, we employ a combination of density-functional theory (DFT) \cite{PhysRev.136.B864,PhysRev.140.A1133}, dynamical mean-field theory (DMFT) \cite{PhysRevLett.62.324,RevModPhys.68.13,kotliar2004strongly,held2007electronic}, and dynamical vertex approximation (D$\Gamma$A) \cite{held2008dynamical,galler2019abinitiodgammaa,RevModPhys.90.025003}  as diagrammatic extensions of DMFT. We show that out-of-plane intercalated hydrogen (H) in La$_2$NiO$_4$ forms a proton that transfers an electron to Ni, converting Ni$^{2+}$ (3$d^8$) to Ni$^+$ (3$d^9$). This creates a single-band structure analogous to that in La$_2$CuO$_4$. Our D$\Gamma$A calculations further predict that this protonated phase, H-La$_2$NiO$_4$, is in close proximity to the infinite-layer nickelate regime and is promising to exhibit superconductivity with a critical temperature above 20\,K. These findings suggest that proton evolution provides a viable path for engineering superconductivity in nickelate systems and opens new avenues for exploring unconventional superconductivity beyond the cuprate family.

\emph{Method---}DFT-level structural relaxations and magnetic and electronic band structure calculations were performed using the \textsc{Vasp}  \cite{kresse1996efficiency,PhysRevB.54.11169} and \textsc{Wien2K} \cite{blaha2001wien2k,Schwarz2002} codes with the Perdew-Burke-Ernzerhof version of the generalized gradient approximation (GGA-PBE) \cite{PhysRevLett.77.3865} and a dense 13$\times$13$\times$5 $k$-mesh for the unit cell of La$_2$NiO$_4$ with and without intercalated H. The interaction parameters for single-band Ni-3$d_{x^2-y^2}$ and full Ni-3$d$ orbitals were obtained from the constrained random phase approximation (cRPA) \cite{PhysRevB.77.085122} implemented in \textsc{VASP} \cite{PhysRevMaterials.9.015001}. 
%The interaction parameters for multi- and single-band DMFT and D$\Gamma$A are $U$(Ni-$d$)=2.92\,eV, $J$(Ni-$d$)=0.62\,eV and $U'$(Ni-$d$)=1.68\,eV, $U(d_{x^2-y^2})$=2.81\,eV.
\emph{Ab initio} molecular dynamics (AIMD) simulations implemented in \textsc{VASP} \cite{PhysRevB.48.13115,PhysRevB.49.14251,PhysRevB.47.558} were performed to study the finite-temperature structural and dynamic properties of hydrided La$_2$NiO$_4$ (H-La$_2$NiO$_4$). The AIMD simulations were conducted in the canonical ensemble using a Nos$\acute{e}$-Hoover thermostat (NVT) to maintain constant temperature. A time step of 1\,fs was used to accurately capture the atomic motions, and each simulation was run for a total of 10$^3$\,fs to ensure adequate statistical sampling of the phase space. A large 2$\times$2$\times$1 supercell of H-La$_2$NiO$_4$ was employed in AIMD to minimize finite-size effects.

Starting points for the DMFT calculations are the \verb|Wien2K| bands of single Ni-3$d_{x^2-y^2}$ and full Ni-3$d$ orbitals around Fermi level  projected onto the maximally localized Wannier functions \cite{PhysRev.52.191,RevModPhys.84.1419} using the \verb|Wien2Wannier|  \cite{mostofi2008wannier90,kunevs2010wien2wannier} interface. The thus obtained DFT-Wannier Hamiltonian is supplemented by a local Kanamori interaction, and we employ the fully localized limit as the double counting correction \cite{PhysRevB.52.R5467,PhysRevB.48.16929}. The resulting many-body Hamiltonian is solved at room temperature (300\,K) within DMFT employing the continuous-time quantum Monte Carlo solver in the hybridization expansions \cite{RevModPhys.83.349} of the \verb|W2dynamics| code \cite{PhysRevB.86.155158,wallerberger2019w2dynamics}. 
Real-frequency spectra are obtained  via analytic continuation using the maximum entropy method (MaxEnt) \cite{PhysRevB.44.6011,PhysRevB.57.10287} as implemented in the \verb|ana_cont| code \cite{kaufmann2023ana_cont}.

The superconducting $T_C$ is calculated with  the D$\Gamma$A \cite{galler2019abinitiodgammaa, held2008dynamical,RevModPhys.90.025003}. First, we compute the particle-particle vertex incorporating spin fluctuations from both the particle-hole and transversal particle-hole channels. Next, we extract the leading superconducting eigenvalues in the particle-particle channel \cite{PhysRevB.99.041115}. This procedure effectively represents the first iteration of the particle-particle channel in a full parquet D$\Gamma$A calculation. For further technical details, see \cite{Kitatani2022}.

\begin{figure*}
\centering
\includegraphics[width=0.9\textwidth]{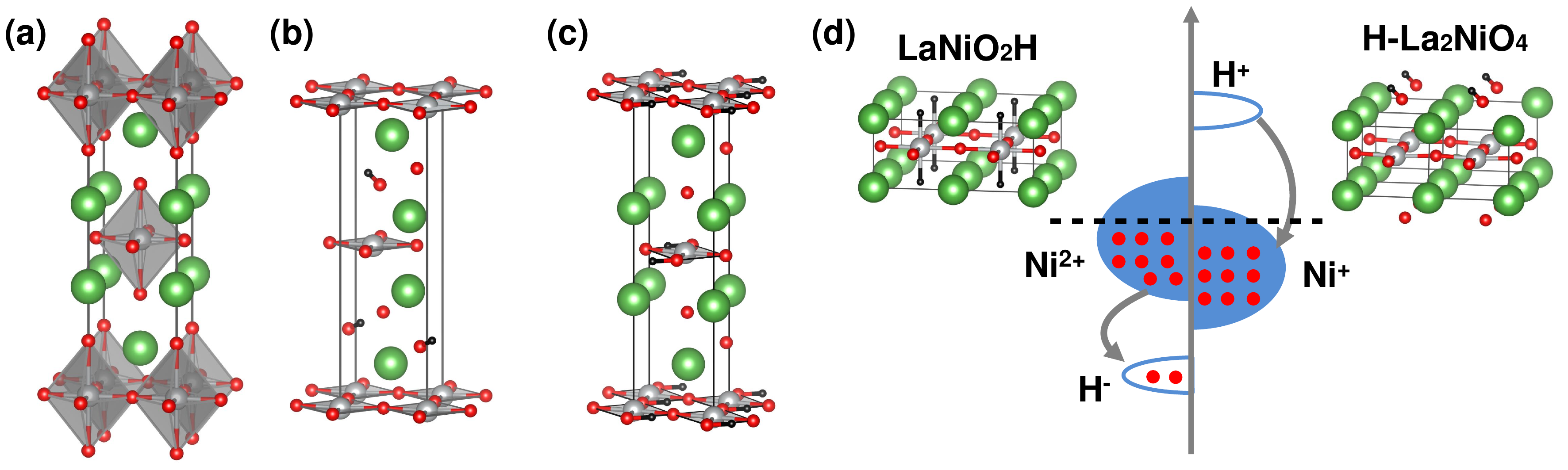}
\caption{Relaxed crystal structures of (a) undoped La$_2$NiO$_4$, (b) H-La$_2$NiO$_4$ with out-of-plane H, (c) H-La$_2$NiO$_4$ with in-plane H; (d) schematic figure of the hydrogen evolution in infinite-layer LaNiO$_2$ and Ruddlesden-Popper (RP) phase La$_2$NiO$_4$. The density of states near Fermi level (dashed line) are from Ni-3$d$ orbitals: in infinite-layer LaNiO$_2$H \cite{PhysRevLett.124.166402} the Ni ions are in a Ni$^{2+}$ valence state, while in the hydride RP phase H-La$_2$NiO$_4$ they are in Ni$^+$ state.}
\label{Fig1_structure}
\end{figure*}

\emph{Structure---}We first investigate the doping position of intercalated H in the Ruddlesden-Popper (RP) phase of La$_2$NiO$_4$. The undoped La$_2$NiO$_4$ (space group $I4/mmm$, No.~139) is illustrated in Fig.~\ref{Fig1_structure}(a). Figures~\ref{Fig1_structure}(b) and \ref{Fig1_structure}(c) present the relaxed structures of H-La$_2$NiO$_4$ with two favorable sites for H intercalation: out-of-plane (OOP) [Fig.~\ref{Fig1_structure}(b)] and in-plane (IP) [Fig.~\ref{Fig1_structure}(c)] position. First-principles DFT structural optimizations reveal that the out-of-plane configuration has a total energy of --111.98\,eV per unit cell [Fig.~\ref{Fig1_structure}(b)] compared to --111.42\,eV [Fig.~\ref{Fig1_structure}(c)] for the in-plane configuration, indicating that the out-of-plane H is more stable by 280\,meV per H atom \footnote{In the present study, we initially designed several possible structures of H-La$_2$NiO$_4$ by intercalating hydrogen into La$_2$NiO$_4$, guided by crystal symmetry analysis. Subsequent nonmagnetic DFT calculations were performed to optimize these structures and determine the ground state of H-La$_2$NiO$_4$. However, a more comprehensive structural search may require exploring a broader range of possible configurations within larger crystal cells. Our primary focus in this research is to investigate the electron doping effect introduced by hydrogen intercalation in La$_2$NiO$_4$}. For details of DFT calculations for other possible structures of H-La$_2$NiO$_4$, see Supplemental Material (SM) \cite{SM} Sec.~I.

\begin{table*}[htbp]
\caption{Crystal lattice constants $a$, $b$, $c$ (\AA),  first, second, third nearest $d_{x^2-y^2}$ intraorbital hopping ($t$, $t'$, $t''$), cRPA computed single-band $U$ (in unit of eV), $U/t$, magnetic ground state and critical temperature $T_C$ (experimental and theoretical values) of the three materials.}
\begin{tabular}{c|c|c|c|c|c|c|c|c|c|c}
\hline
\hline
Materials & $a$ (\AA) & $c$ (\AA) & $t$  & $t'/t$ & $t''/t$ & $U$ & $U/t$ & Magnetism & $T_C$ (experimental) & $T_C$ (theory) \\
\hline
La$_2$CuO$_4$ (this work) & 3.814 & 13.145 & -0.444  & -0.055 & 0.081 & 2.68 & 6.0 & $G$-AFM & 30\,K \cite{bednorz1986possible} & --- \\\hline
\hline
(Nd,La)NiO$_2$ \cite{kitatani2020nickelate,PhysRevLett.124.166402} & 3.890 & 3.337 & -0.395  & -0.240 & 0.118 & 2.60 & 6.5 & $G$-AFM & 20\,K \cite{lee2023linear} & 36\,K \cite{kitatani2020nickelate} \\
\hline
H-La$_2$NiO$_4$ (this work) & 3.913 & 13.712 & -0.356  & -0.234 & 0.105 & 2.81 & 7.9 & $G$-AFM & --- & 20\,K\\
\hline
\end{tabular}
\label{Tab1}
\end{table*}

\begin{figure*}
\centering
\includegraphics[width=0.85\textwidth]{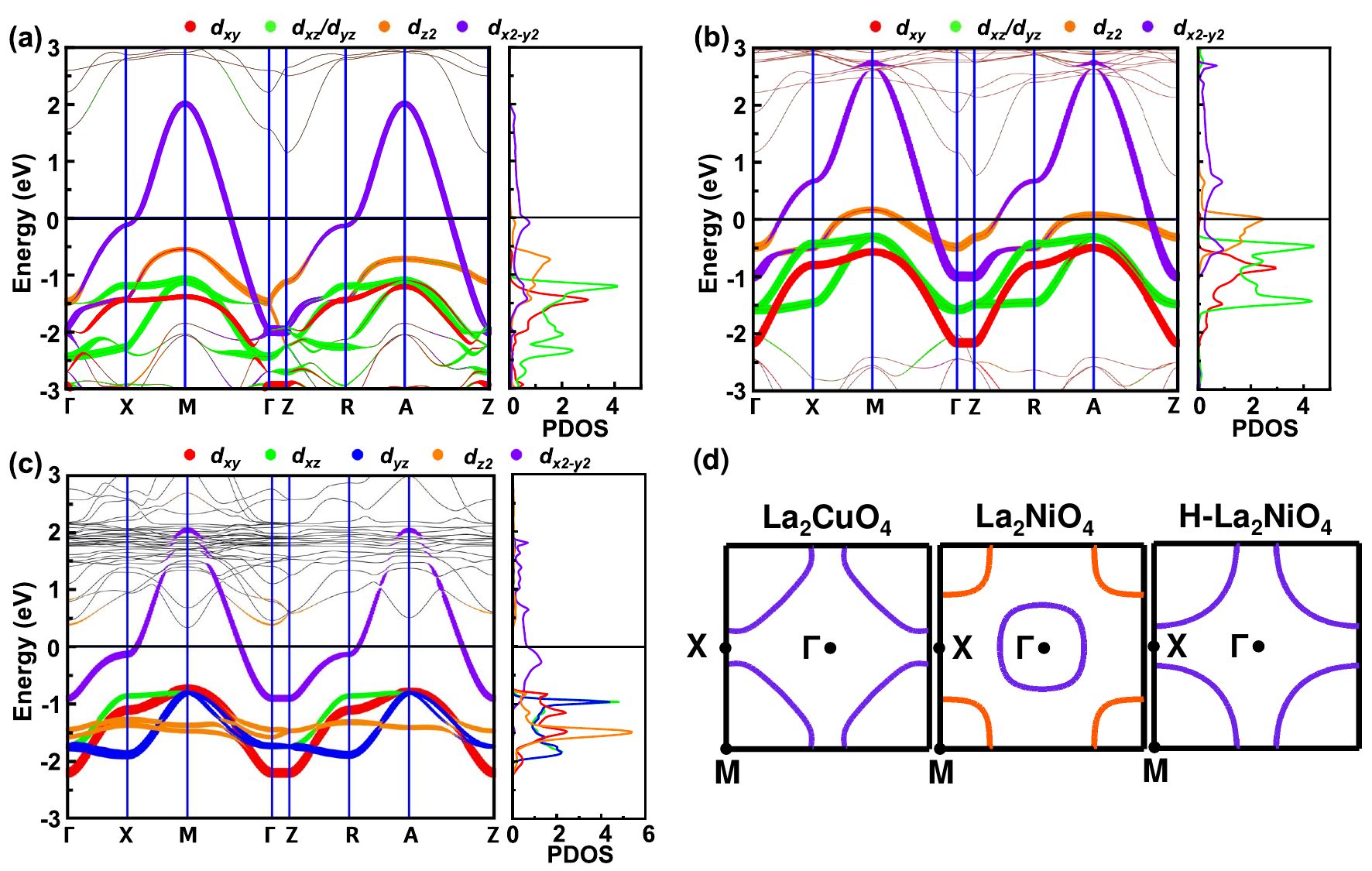}
\caption{DFT band structures of (a) La$_2$CuO$_4$, (b) La$_2$NiO$_4$, (c) H-La$_2$NiO$_4$, and (d) corresponding Fermi surfaces at $k_z$=0. The band characters of Cu- and Ni-3$d$ orbitals in (a)--(c) are labeled by different colors.}
\label{Fig2_DFT}
\end{figure*}

Since DFT calculations are performed at zero temperature, we validated the stability at finite temperature using AIMD (embedded in \textsc{VASP} \cite{PhysRevB.48.13115,PhysRevB.49.14251,PhysRevB.47.558}). Figure~\ref{Fig2_DFT}  \textcolor{blue}{in SM \cite{SM} Sec.~II} shows the crystal at 300\,K. Notably, DFT optimization and MD simulations at 300\,K yield H-O bond lengths of 0.99\,\AA~in DFT and 0.99--1.05\,\AA~in MD at 300\,K \cite{SM}, respectively, consistent with typical H-O bond lengths. Thus, the out-of-plane position doped H forms a stable, negatively charged hydroxide (OH)$^-$ with connection to the out-of-plane O.
In previous work on infinite-layer nickelate LaNiO$_2$ with H \cite{PhysRevLett.102.226401,PhysRevB.107.165116,PhysRevB.108.165145}, DFT and DMFT calculations revealed that H ions form a $\delta$-type bond with Ni via the Ni-3$d_{z^2}$ and H-1$s$ orbital, resulting in an H$^-$  valence state and the formation of a Ni$^{2+}$  (3$d^8$) configuration [see Fig.~\ref{Fig1_structure}(d)]. For out-of-plane H-La$_2$NiO$_4$ [Fig.~\ref{Fig1_structure}(b)], which has lower energy in DFT, an opening question is whether electron doping can be achieved by intercalating H.

\emph{DFT bands---}To address this question, we first performed non-spin-polarized DFT calculations for the RP-phase La$_2$CuO$_4$---the parent compound of single-band cuprate superconductor---and La$_2$NiO$_4$, a two-band AFM Mott insulator. Figure~\ref{Fig2_DFT}(a) shows that La$_2$CuO$_4$ features a single-band Fermi surface.
In contrast, La$_2$NiO$_4$ exhibits a two-band Fermi surface, with Ni-3$d_{z^2}$ and 3$d_{x^2-y^2}$ orbitals contributing a holelike Fermi surface 
and an electron-type pocket, respectively. In H-La$_2$NiO$_4$, however, the Fermi surface displays a single-band $d_{x^2-y^2}$ feature as the $d_{z^2}$ band shifts down compared to its position in La$_2$NiO$_4$ [Fig.~\ref{Fig2_DFT}(b) and \ref{Fig2_DFT} (c)]. This indicates that the H ion adopts an H$^+$ valence state. More specifically, in H-La$_2$NiO$_4$, H forms hydroxide (OH)$^-$ with O [Fig.~\ref{Fig1_structure}(d)], resulting in a 3$d^9$ (Ni$^+$) configuration and consequently a characteristic single-band  Fermi surface [Fig.~\ref{Fig2_DFT}(d)], remarkably similar to that of the cuprate La$_2$CuO$_4$ and infinite-layer LaNiO$_2$. SM \cite{SM}  Secs. III--V demonstrate that this single-Ni-band picture  is robust against both the position of topotactic H and strain. Remarkably, the latter can lead to a self-doping akin to infinite-layer nickelates.

Additionally, we calculated the electron hopping parameters of H-La$_2$NiO$_4$ using the Wannier projection method \cite{PhysRevB.56.12847,PhysRevB.65.035109,RevModPhys.84.1419}, including the first ($t$), second ($t'$), and third ($t''$) nearest-neighbor hoppings, as well as the ratios $t'$/$t$ and $t''$/$t$. Our findings indicate that these hopping parameters in H-La$_2$NiO$_4$ closely resemble those found in infinite-layer (La,Nd)NiO$_2$ (Table~\ref{Tab1}), such as $t'$/$t$ $\sim$ --0.24. Previous studies \cite{PhysRevLett.130.166002} have demonstrated that these hopping parameters play a crucial role in determining the superconducting properties within the two-dimensional Hubbard model framework, especially for nickelates and palladates. They influence not only the emergence of superconductivity but also the maximum superconducting transition temperature ($T_C$) achievable in these materials. The similarity in hopping parameters suggests that H-La$_2$NiO$_4$ may exhibit comparable superconducting characteristics to (La,Nd)NiO$_2$.
Moreover, our DFT+$U$ magnetic order calculations indicate that the ground state of H-La$_2$NiO$_4$ is an antiferromagnetic Mott insulator (see SM \cite{SM} Sec.~VI). This finding aligns with the behavior observed (experimentally) in parent compounds of cuprate superconductors, such as La$_2$CuO$_4$, further highlighting the parallels between these material systems.

As a consequence, both DFT band calculations and Wannier projections reveal notable electronic structure similarities between H-La$_2$NiO$_4$, LaNiO$_2$ and La$_2$CuO$_4$. These materials exhibit features consistent with the one-band Hubbard model, which has been pivotal in understanding high-temperature superconductivity. This resemblance again underscores the potential of H-La$_2$NiO$_4$ as a candidate for high-temperature superconductivity, akin to its nickelate counterparts.

\emph{Correlations---}DFT calculations often underestimate electronic correlations, particularly the dynamical correlation effects within 3$d$ orbitals. Therefore, we performed DMFT calculations to examine the electronic structure near the Fermi level in greater detail. \textsc{VASP}-cRPA calculations for H-La$_2$NiO$_4$ were performed using a multiband and single-band model, respectively: for the full-$d$ model, we obtain an
intraorbital $U$=2.92\,eV, 
Hund's $J$=0.62\,eV, and interorbital $U'$=$U$-2$J$=1.68\,eV; for the single-band model the Hubbard is $U$=2.8\,eV, leading to $U$/$t$$\sim$7.9 ($t$: the nearest intraorbital   hoppings of this single $d_{x^2-y^2}$ orbital are summarized in  Table~\ref{Tab1}). 

\begin{figure*}
\centering
\includegraphics[width=0.8\textwidth]{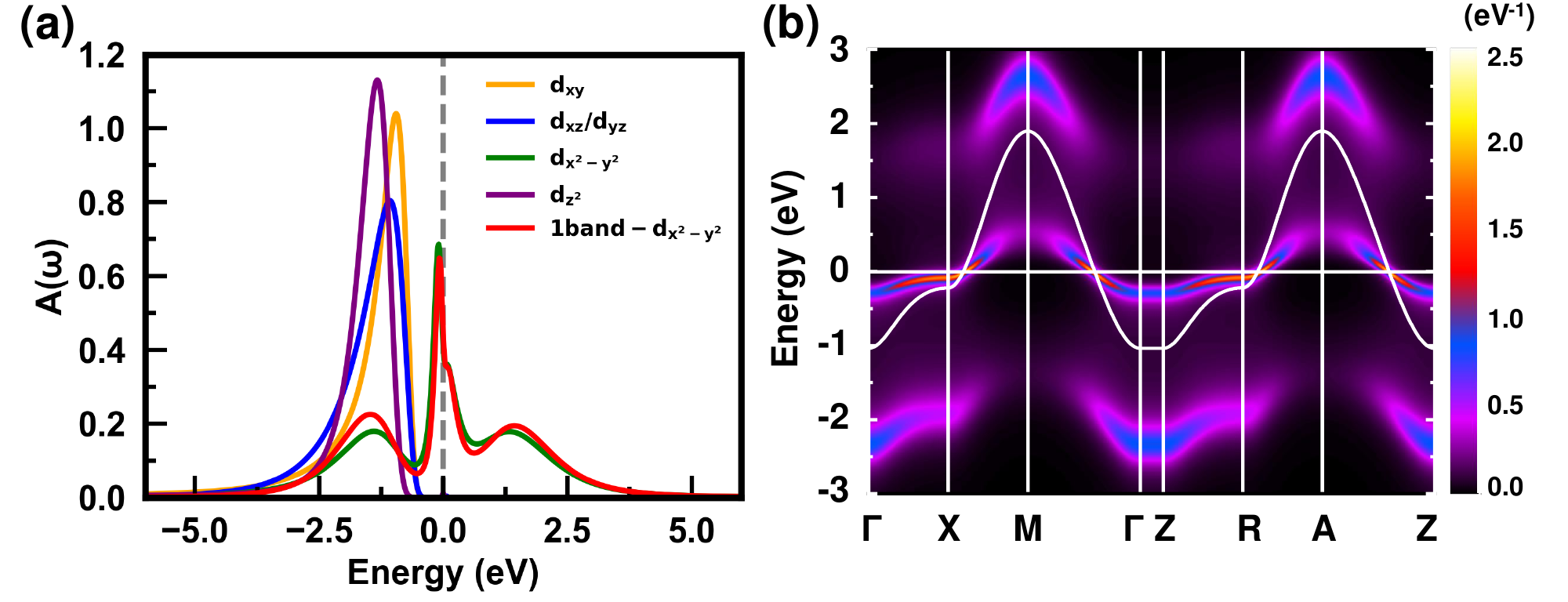}
\caption{(a) Multi- and single-orbital $k$-integrated DMFT spectral functions $A(\omega)$. (b) Single-orbital $k$-resolved spectral function $A$($\omega$,$k$) of H-La$_2$NiO$_4$. For DMFT spectral functions of the in-plane phase of H-La$_2$NiO$_4$ and under different strains see SM \cite{SM} Sec. V.}
\label{Fig3_DMFT}
\end{figure*}

Figure \ref{Fig3_DMFT}(a) compares the $k$-integrated  DMFT spectral functions 
$A$($\omega$) for multi- and single-orbital calculations. In the multiorbital $A$($\omega$), the $d_{z^2}$ and $t_{2g}$ orbitals are fully occupied, while the $d_{x^2-y^2}$ orbital is half-filled and exhibits significant renormalization, with an effective mass $m^*/m_e$ of approximately 3.3. This value is close to that reported for infinite-layer nickelates (4.4) \cite{kitatani2020nickelate} and for some cuprates, it also ranges from 3 to 4 \cite{PhysRevLett.62.2317,PhysRevB.72.060511,PhysRevB.105.085110}. Notably, the single-band 
$A$($\omega$) shown in Fig.~\ref{Fig3_DMFT}(a) exhibits remarkable similarity to the full-$d$ multibands $A$($\omega$), which strongly supports the validity of a minimal low-energy model based solely on the $d_{x^2-y^2}$ orbital in H-La$_2$NiO$_4$. Additionally, we computed the $k$-resolved spectral function 
$A$($k$,$\omega$) for the $d_{x^2-y^2}$ orbital [Fig.~\ref{Fig3_DMFT}(b)], revealing that the Fermi surface is entirely derived from this band, with the upper and lower Hubbard bands appearing at approximately 2\,eV and --2\,eV, respectively. 
Notably, at $X$ and $R$ points, the Ni-3$d_{x^2-y^2}$ band exhibits nearly identical energies, indicating its two-dimensional (2D) character. This arises from the substantial spatial separation of Ni ions along the $z$ axis, which minimizes interlayer interactions/hopping and reinforces its 2D character.
DFT+$U$ calculations for a 2$\times$2$\times$1 supercell of H-La$_2$NiO$_4$ indicate the ground state is a $G$-type AFM Mott-insulating state (see SM \cite{SM} Sec.~VI).

\begin{figure*}
\centering
\includegraphics[width=0.9\textwidth]{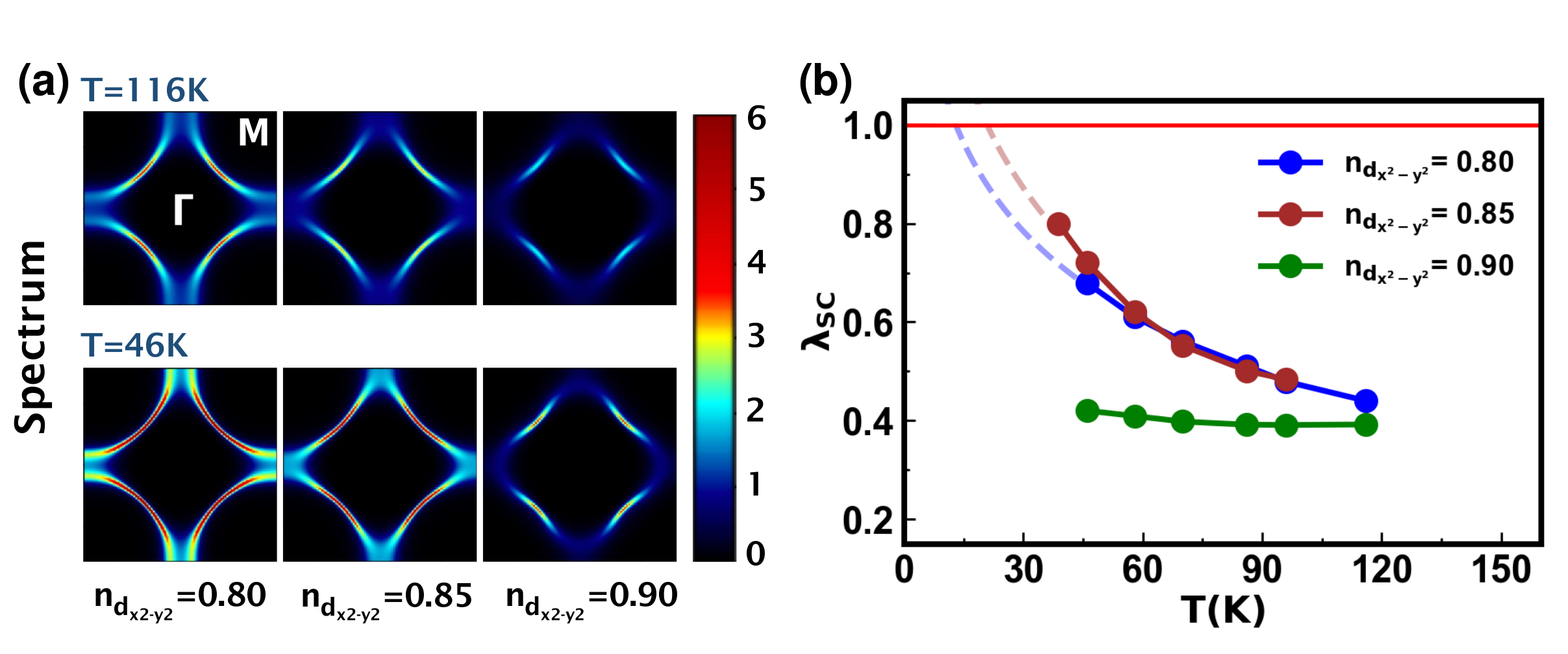}
\caption{(a) D$\Gamma$A $k$-resolved spectral function of H-La$_2$NiO$_4$ at the Fermi energy for $T$=116\,K (top) and $T$=46\,K (bottom), shown for three doping levels $n_{d_{x^2-y^2}}$ of the Ni-3$d_{x^2-y^2}$ band. (b) Leading superconducting $\lambda_{SC}$  eigenvalue (here $d$-wave) as a function of temperature $T$ using the form $\lambda \approx a - b$ ln($T$) as in previous publications \cite{kitatani2020nickelate,Kitatani2022};  $\lambda_{SC}$$\rightarrow$1 (intersection with the red line) signals the onset of superconducting ($T_C$). The dashed lines indicate the fitting of $\lambda_{SC}$ at $n_{d_{x^2-y^2}}$=0.80 and 0.85, respectively.
For $n_{d_{x^2-y^2}}$=0.80 and $n_{d_{x^2-y^2}}$=0.85, $T_C=13.04$\,K and 20.87\,K,  respectively.}
\label{Fig4_DGA}
\end{figure*}

\emph{Estimating $T_C$---}The above DFT and DMFT results demonstrate the effectiveness of a minimal single-band model and its essentially 2D character. Similar to the $d_{x^2-y^2}$ Hubbard model for cuprates and nickelates, this single band captures the low-energy electronic structure and prospectively the superconducting properties. To further investigate superconductivity, more sophisticated many-body methods than DMFT, such as D$\Gamma$A method \cite{RevModPhys.90.025003}, must be employed. D$\Gamma$A starts from a local irreducible vertex (depending on three frequencies and obtained at DMFT convergence) and builds nonlocal vertices and self-energies through ladder or parquet diagrams. The starting vertex in D$\Gamma$A is nonperturbative, unlike in RPA or FLEX. This approach incorporates both the local DMFT correlations and the nonlocal spin, charge, and superconducting fluctuations, enabling the calculation of quantum critical exponents, pseudogaps, and superconducting properties in the one-band Hubbard model. Previous applications \cite{PhysRevB.99.041115,kitatani2020nickelate,PhysRevLett.130.166002,PhysRevMaterials.6.L091801,PhysRevB.109.235126} have proven its power in predicting superconducting materials materials and estimating critical temperatures in cuprates and nickelates.

Figure~\ref{Fig4_DGA}(a) displays the D$\Gamma$A Fermi surface at various hole dopings (10\%, 15\% and 20\%) and temperatures (46\,K and 116\,K), based on the constructed 2D $d_{x^2-y^2}$ Hubbard model with the hopping parameters from Table~\ref{Tab1}. The Fermi surface is obtained from the lowest Matsubara frequency. Please note that the interlayer ($z$-direction) hopping between $d_{x^2-y^2}$ orbital is justified by the spacing layers, which makes H-La$_2$NiO$_4$ an even more ideal two-dimensional system  than infinite-layer Nd(La)NiO$_2$ (which has a $z$ hopping of $\sim$34\,meV compared to $\sim$2\,meV in H-La$_2$NiO$_4$). For a doping level $n$=0.85 (corresponding to H-La$_{1.85}$Sr$_{0.15}$NiO$_4$), a well-defined holelike Fermi surface can be observed in Fig.~\ref{Fig4_DGA}(a). In contrast, at $n$=0.9 (H-La$_{1.9}$Sr$_{0.1}$NiO$_4$), strong AFM spin fluctuations induce significant scattering at the antinodal point ${\mathbf k}$=($\pi$,0). This signals the vicinity of the pseudogap regime.
At the lower temperature of 46\,K, the bands become more well-defined across the Brillouin zone, indicating a suppression of the $d_{x^2-y^2}$ hole contributions to, e.g., Hall coefficient at higher temperatures. 
%\khc{Say how A is calcualted: from maxent? $\tau=\beta/2$}

The superconducting critical temperature $T_C$ can be determined from the divergence of the superconducting susceptibility $\chi$, or equivalently, the leading superconducting eigenvalue $\lambda_{SC}$. In matrix notation \cite{PhysRevB.99.041115}, $\chi$=$\chi_0$/[1-$\Gamma_{pp}$$\chi_0$], where $\chi_0$ is the bare superconducting susceptibility and $\Gamma_{pp}$ is the irreducible vertex in the particle-particle channel calculated by D$\Gamma$A. When the leading eigenvalue $\lambda_{SC}$ of $\Gamma_{pp}$$\chi_0$ approaches 1, the superconducting susceptibility diverges. Figure~\ref{Fig4_DGA}(b) plots $\lambda_{SC}$ versus temperature at different fillings of the Ni-$d_{x^2-y^2}$ orbital. It shows that $\lambda_{SC}$ approaches 1 at low temperatures for $n$=0.80 and $n$=0.85, whereas for $n$=0.90, there seems to be no superconductivity. Specifically, the superconducting state is absent for
$n$=0.9 down to the lowest measured temperature, and for $n$=0.8, it is only observed at T$<$13\,K. This suggests a superconducting dome  within the range 0.90$>$n$\geq$0.80. The momentum structure
of the leading eigenvector (not shown) indicates $d$-wave superconductivity.

\emph{Conclusion---}The recent discovery of infinite-layer nickelate superconductors has revealed a wide range of crystal structures, chemical compositions, and potential superconducting mechanisms in different nickelates. In contrast, although La$_2$NiO$_4$ has been extensively studied and exhibits a variety of novel quantum states, superconductivity has not been found yet experimentally. Here, we investigated the structural, thermal, magnetic, electronic, and superconducting properties of H-La$_2$NiO$_4$ where hydrogen is topotactically intercalated, using state-of-the-art theoretical methods. These methods include D$\Gamma$A, which is well-suited for calculating superconducting temperatures in two-dimensional strongly correlated oxides such as cuprates and nickelates.
Unlike in LaNiO$_2$H,  it acts as an electronic donor in H-La$_2$NiO$_4$.  This leads to a  Ni$^+$ valence with an electronic configuration similar to Cu$^{2+}$. Also the in-plane AFM Mott insulating state for full hydrogen intercalation resembles that of cuprate superconductors such as La$_2$CuO$_4$. With an optimal hole doping of approximately 15\%, our simulations predict that H-La$_2$NiO$_4$ exhibits a $d$-wave superconducting state with $T_C$ up to 20\,K, akin to its infinite-layer nickelate analogs.
Against the background of recent experimental progress to add hydrogen via ionic liquid gating in perovskite and RP-oxides, the experimental synthesis of the suggested H-La$_2$NiO$_4$ appears possible and promising.

\emph{Acknowledgments---}We thank Prof.~Pu Yu, Viktor Christiansson and Eric Jacob for helpful insightful discussions.
L.~S.~acknowledges support from the National Natural Science Foundation of China (Grant No.~12422407).
L.~S. and K.~H.~acknowledge funding through the Austrian Science Funds (FWF) Project Grant DOI 10.55776/I5398.
W.-F.~W.~acknowledges the funding from the China Scholarship Council (CSC).
Calculations have been done on the the National Supercomputing Center (Xi'an) in Northwest University and the Vienna Scientific Cluster (VSC).

%\bibliographystyle{unsrtnat}
%\bibliography{main}

\end{document}